\documentclass[prc,twocolumn,showpacs,preprintnumbers,
               amsmath,amssymb,
               superscriptaddress]{revtex4}
\usepackage{graphicx} % Include figure files
\usepackage{dcolumn}  % Align table columns on decimal point
\usepackage{bm}       % bold math

\newcommand{\Vvec}{\boldsymbol V}
\newcommand{\Fvec}{\boldsymbol F}
\newcommand{\Ivec}{\boldsymbol I}
\newcommand{\uvec}{\boldsymbol u}

\begin{document}

\title{Broyden's Method in Nuclear Structure Calculations}

\author{Andrzej Baran}
 \affiliation{Department of Physics \&
  Astronomy, University of Tennessee, Knoxville, Tennessee 37996, USA}
\affiliation{Physics Division, Oak Ridge National Laboratory, P.O. Box
  2008, Oak Ridge, Tennessee 37831, USA}
 \affiliation{Institute of Physics, University of M. Curie-Sklodowska,
  ul. Radziszewskiego 10, 20-031 Lublin, Poland}

\author{Aurel Bulgac}
\affiliation{Department of Physics, University of Washington, Seattle,
  WA 98195-1560}
  
\author{Michael McNeil Forbes}
\affiliation{Department of Physics, University of Washington, Seattle,
  WA 98195-1560}
  
\author{Gaute Hagen}
\affiliation{Physics Division, Oak Ridge National Laboratory, P.O. Box
  2008, Oak Ridge, Tennessee 37831, USA}

 \author{Witold Nazarewicz} 
\affiliation{Department of Physics \&
  Astronomy, University of Tennessee, Knoxville, Tennessee 37996, USA}
\affiliation{Physics Division, Oak Ridge National Laboratory, P.O. Box
  2008, Oak Ridge, Tennessee 37831, USA}
\affiliation{Institute of
  Theoretical Physics, Warsaw University, ul. Ho\.za 69, 00-681
  Warsaw, Poland}
\affiliation{School of Engineering and Science,
University of the West of Scotland,
Paisley  PA1 2BE, UK} 

\author{Nicolas Schunck}
\affiliation{Department of Physics \&
  Astronomy, University of Tennessee, Knoxville, Tennessee 37996, USA}
\affiliation{Physics Division, Oak Ridge National Laboratory, P.O. Box
  2008, Oak Ridge, Tennessee 37831, USA}
  
\author{Mario V.~Stoitsov}
\affiliation{Department of Physics \&
  Astronomy, University of Tennessee, Knoxville, Tennessee 37996, USA}
\affiliation{Physics Division, Oak Ridge National Laboratory, P.O. Box
  2008, Oak Ridge, Tennessee 37831, USA}
\affiliation{Institute of Nuclear Research and Nuclear Energy,
  Bulgarian Academy of Sciences, Sofia, Bulgaria}

\date{\today}

\newcommand{\BdG}{\textsc{b}{\footnotesize d}\textsc{g}}
\newcommand{\PWscf}{\textsc{pw}scf}
\newcommand{\DFT}{\textsc{dft}}
\newcommand{\SLDA}{\textsc{slda}}
\newcommand{\HO}{\textsc{ho}}
\newcommand{\THO}{\textsc{tho}}
\newcommand{\HF}{\textsc{hf}}
\newcommand{\LN}{\textsc{ln}}
\newcommand{\HFB}{\textsc{hfb}}
\newcommand{\HFODD}{\textsc{hfodd}}
\newcommand{\HFBTHO}{\textsc{hfbtho}}
\newcommand{\EDF}{\textsc{edf}}
\newcommand{\DVR}{\textsc{dvr}}
\newcommand{\CC}{\textsc{cc}}
\newcommand{\CCSD}{\textsc{ccsd}}
\newcommand{\BLAS}{\textsc{blas}}
\newcommand{\LAPACK}{\textsc{lapack}}
\newcommand{\GNU}{\textsc{gnu}}
\newcommand{\exclude}[1]{}
\newcommand{\mat}[1]{\mathbf{#1}}
\newcommand{\ket}[1]{|#1\rangle}
\newcommand{\bra}[1]{\langle#1|}
\newcommand{\norm}[1]{\lVert{#1}\rVert}
\newcommand{\braket}[1]{\mathinner{\langle{#1}\rangle}}{\catcode`\|=\active
  \gdef\Braket#1{\left<\mathcode`\|"8000\let|\bravert {#1}\right>}}
\newcommand{\bravert}{\egroup\,\vrule\,\bgroup}

\begin{abstract}
Broyden's method, widely used in quantum chemistry
electronic-structure calculations
for the numerical solution of nonlinear equations 
in many variables, is applied in the context of the nuclear many-body problem.
Examples include  the unitary gas problem, the nuclear density
functional theory with Skyrme functionals,  and
the nuclear  coupled-cluster theory.
The stability of the method, its ease of use, and its rapid convergence rates
make Broyden's method a tool of choice for large-scale nuclear structure
calculations.
\end{abstract}


\maketitle

\section{Introduction}
\label{sec:introduction}

The nuclear many-body problem is undergoing a renaissance.
 Hand in hand with the 
experimental developments in the science of rare isotopes, 
a qualitative change in theoretical modeling
of the nucleus is taking place. 
 Developments of powerful conceptual, analytic, algorithmic,
and computational tools enable scientists to probe the inner
workings of nuclei with far greater precision than previously possible.
These new tools, including terascale computer platforms,
make researchers optimistic that the goal of developing
a comprehensive, quantitative, and predictive theory of the nucleus and
nucleonic matter is  achievable \cite{RIAtheory,unedf}.

The research challenges faced by the nuclear structure 
community are often interdisciplinary. Indeed, the quantum many-body
problem represents one of 
the great intellectual and numerical challenges for
nuclear and hadronic physics, quantum chemistry, 
atomic and condensed matter physics,
and materials sciences. Often, to solve  problems in many-body science,
close interactions of physicists with computer scientists and
mathematicians are required. 

A theoretical framework aiming at the microscopic description of 
many-body  systems
and capable of extrapolating into  unknown  regions must fulfill
several requirements. Namely,  (i) it must be general enough to be confidently
applied to unknown species  whose properties are largely unknown;
(ii) it should be capable of handling symmetry-breaking effects inherent
to finite systems; (iii)
it should be able to 
describe both finite systems and bulk matter; and (iv)
in addition to observables, the method should provide associated error bars.
These requirements are met by the Density Functional Theory (\DFT)
in the formulation of Kohn and Sham \cite{KS}.

The generalization of the \DFT\ to the case of fermionic
pairing was formulated for electronic superconductors  
\cite{DFTpair}. The resulting 
Hartree-Fock-Bogoliubov (\HFB) or 
Bogoliubov de-Gennes (\BdG) equations can be
viewed as the generalized Kohn-Sham equations of the
standard \DFT. 

The nuclear \DFT\  provides a reliable method for
calculating properties of nuclei across the whole nuclear mass chart. 
The main ingredient of the nuclear \DFT\ \cite{DFTrev}
is the energy density functional that depends on proton and neutron 
densities and currents representing distributions of nucleonic matter, spins,
momentum, and kinetic energy, as well as their derivatives (gradient terms).
If pairing correlations are present, the particle densities are augmented by 
pairing densities representing correlated nucleonic pairs \cite{LDA}.
Initially, attempts to build
effective energy functionals were rooted in the empirical zero-range
Skyrme interaction treated within the Hartree-Fock (\HF) or
\HFB\ approximation. Following the success of
\DFT\ in atomic and molecular physics and, more recently, in the physics
of ultra-cold atoms, it was realized that the interaction could
be secondary to the functional, and recent progress has been directed
towards extensions of the energy functionals to less traditional forms.

In parallel, much effort is put into trying to connect the
nuclear \DFT\  to more fundamental many-body theory based on
inter-nucleon interactions.  In order to extend the \emph{ab-initio} program to the
medium-mass region of the nuclear chart, a method which scales softly
with system size and has a controllable error estimate is called
for. The Coupled-Cluster (\CC) theory is ideal in this respect. It scales
polynomial with system size and it is {\it size extensive}, meaning
that the energy scales correctly with increasing system
size. 

This work  represents  a collaborative effort performed
 under  the Universal Nuclear
Energy Density Functional ({\sc unedf}) project \cite{unedf}. The goal
of {\sc unedf} is to develop a new-generation
theoretical framework that will  describe all nuclei and
their reactions  by making  use of the massive computer power.
One of the objectives  of {\sc unedf} is to develop 
computational infrastructure for leadership-class computers;
hence the algorithmic and computational focus of this paper.

The common feature of  \DFT\  and \CC\ is to
solve the quantum many-body problem for interacting fermions 
through self-consistent convergence schemes. The computational cost of
these iterative procedures can become very expensive, especially when
the size of the model space  or the number of nuclei
processed simultaneously is huge. The
advent of teraflop computers  make such
large-scale calculations feasible, but in order to take full advantage
of unique resources,  better
convergence algorithms are called for. 
The aim of the present paper is to report the  implementation of
Broyden's method \cite{broyden,mbm},
a quasi-Newton algorithm to solve large sets of non-linear equations,
 in nuclear structure calculations. 
The \HFB\ (or \BdG) equations
-- the Kohn-Sham equations of the superfluid \DFT\ --
represent a coupled system of non-linear equations for 
densities or mean fields. The non-linearity enters through the 
self-consistency, i.e., the dependence of  mean fields on densities.
The \CC\ equations have in fact  a very similar structure. Here,
the non-linearity is hidden in the \CC\ intermediates which depend on the
particle-hole amplitudes.
 
Self-consistent formulations of these approaches usually diverge when using
straight iterations. The simplest method commonly used  to avoid 
divergences is the so-called linear mixing, in which  input and
output at a given iteration of the self-consistent process are mixed
to provide input for the next step. We show that
Broyden's method leads to much improved convergence rates.
In many cases, the number of iterations required to reach convergence is
lower by a factor of 3 to 10 with respect to linear-mixing iterative
schemes.

The paper is organized
as follows. Broyden's method is outlined in
Sec.~\ref{sec:broyd-mixing-meth}, both in its standard version 
(Sec.~\ref{SBM}) that applies to a small number of unknowns, and
in a modified version (Sec.~\ref{MBM}) which is used to crunch
 very large 
sets of nonlinear equations. The examples of implementations come next.
Section~\ref{sec:unitary-fermi-gas} discusses the performance of the standard
Broyden's method for 
the case of the unitary gas of two fermionic superfluids in spherical geometry.
In the following examples the modified Broyden's method 
has been  used, as the size of the problem
is intractable with the standard approach. The
general case of the
symmetry-unrestricted Skyrme-{\DFT} theory
is discussed in Sec.~\ref{sec:symm-unconstr-hfb}.
Section~\ref{sec:symm-constr-hfb} illustrates the particular
case of an axially-symmetric Skyrme-{\DFT} problem and discusses
the ability of the method to converge to the local energy extrema
(i.e, minima, maxima, and saddle points).
The performance of the method in  {\em ab-initio} coupled-cluster 
calculations, where the number of unknowns can be as large as $10^8$, 
is presented in  Sec.~\ref{sec:coupl-clust-theory}.
Finally, the conclusions of our work
are contained in Sec.~\ref{conclusions}.

\section{Broyden's Mixing Method}
\label{sec:broyd-mixing-meth}
When solving a system of self-consistent equations, one usually begins
with a set of initial conditions (in nuclear structure:
single-particle wave functions, densities, fields, etc.) that linearize
the problem.  These initial conditions can be represented
formally by an $N$-dimensional vector $\Vvec_{\text{in}}^{(0)}$.  Solving
the self-consistent equations with these initial conditions defines a
new vector
\begin{equation}
  \Vvec^{(m)}_{\text{out}} = \Ivec\bigl(\Vvec_{\text{in}}^{(m)}\bigr)
\end{equation}
where $\Ivec(\Vvec)$ is a function of the initial conditions.  The
self-consistency condition requires that the solution $\Vvec^{*}$ be a
fixed-point of this iteration: $\Ivec(\Vvec^{*}) = \Vvec^{*}$.

The goal of a fixed-point iteration algorithm is to construct a new guess
$\Vvec_{\text{in}}^{(m+1)}$ in such a way that the iteration converges,
\begin{equation}
  \label{FM}
  \Fvec^{(m)} \equiv \Vvec_{\text{out}}^{(m)} - \Vvec_{\text{in}}^{(m)} \rightarrow 0,
\end{equation}
within the required tolerance.
Using the previous output as the next input, $\Vvec_{\text{in}}^{(m+1)} =
\Vvec_{\text{out}}^{(m)}$, does not guarantee convergence.  If
the resulting divergence exhibits oscillatory behavior,  
the usual ansatz is to mix
the input at iteration $m$ to define the input at
iteration $m+1$:
\begin{subequations}
  \label{LM}
  \begin{align}
    \Vvec_{\text{in}}^{(m+1)} &= \alpha \Vvec_{\text{out}}^{(m)} + 
    (1-\alpha) \Vvec_{\text{in}}^{(m)}  \\
    & = \Vvec_{\text{in}}^{(m)} + \alpha \Fvec^{(m)}.
  \end{align}
\end{subequations}
This {\it simple mixing} slows down the iteration, and with 
a suitable choice of the mixing parameter  $\alpha \in [0,1]$,
convergence may be achieved. However, even if one can find  $\alpha$
that leads to a convergent solution, there are many instances where the
convergence is very slow. In large-scale calculations involving huge
numbers of cases, the running time to find poorly converging
solutions may become prohibitive.

If the vector field $\Ivec(\Vvec)$ is differentiable,  one can use
derivative information to improve convergence.  The idea is to regard
the self-consistent condition $\Fvec^{(m)}=~0$ as a set of non-linear
equations, and to approach the problem from the point of view of
finding the root: $\Fvec(\Vvec^{*}) = \Ivec(\Vvec^{*}) - \Vvec^{*} = 0$.  
This set of non-linear equations can be solved  using
the multidimensional Newton-Raphson method, in which the next guess can be
approximated by
\begin{equation}
  \label{BM0}
  \Vvec_{\text{in}}^{(m+1)} = \Vvec_{\text{in}}^{(m)} - \mat{B}^{(m)}\Fvec^{(m)},
\end{equation}
where $\mat{B}^{(m)}=(\mat{J}^{(m)})^{-1}$ and $J^{(m)}_{jk}
= \partial F_{j}^{(m)}(\Vvec)/\partial V_{k}$ is the Jacobian matrix of the
non-linear equations at $\Vvec = \Vvec_{\text{in}}^{(m)}$.  For sufficiently
smooth functions, this method has a quadratic convergence. However, 
this approach is very expensive as it  needs the 
explicit derivative evaluation.

\subsection{Standard Broyden's Method}\label{SBM}

In many situations, the quick evaluation of the inverse Jacobian is
not possible. In such cases one can often
achieve superlinear convergence by using a multidimensional
generalization of the secant method, whereby an approximate matrix
 $\mat{B}^{(m)}$ is computed using a secant approximation based on
the information obtained in previous iterations.  
This method is usually referred to as the quasi-Newton method.
The difficulty in higher dimensions
is that the secant update is not unique, and na\"\i{}ve
implementations fail to converge.  Broyden~\cite{broyden} suggested 
using the Sherman-Morrison formula to form an 
update  to the inverse of the   Jacobian that has good
convergence properties:
\begin{equation}
  \mat{B}^{(m+1)} = \mat{B}^{(m)} + 
  \frac
  {\ket{\delta \Vvec_{\text{in}}}-\mat{B}^{(m)}\ket{\delta\Fvec}}
  {\bra{\delta\Vvec_{\text{in}}}\mat{B}^{(m)}\ket{\delta\Fvec}}
  \bra{\delta \Vvec_{\text{in}}}\mat{B}^{(m)},
\end{equation}
where
\begin{align*}
  \ket{\delta\Vvec_{\text{in}}} &= \Vvec_{\text{in}}^{(m+1)} - \Vvec_{\text{in}}^{(m)}, \\
  \ket{\delta\Fvec}  &= \Fvec^{(m+1)} - \Fvec^{(m)}.
\end{align*}
One can start the iterative process with the initial guess $\mat{B}^{(0)} = \alpha\mat{1}$,
as this is equivalent to the simple mixing~(\ref{LM}).  At any given
step, one may also check the
full quasi-Newton step~(\ref{BM0}): if it produces larger residuals
$|\Fvec^{(m+1)}|$ or increases the energy of the solution, then one
might revert to  simple  mixing (\ref{LM}).  This can be
especially important if the function $\Fvec(\Vvec)$ varies significantly with
the step size because the secant approximation may then
be extremely poor.  More sophisticated, but globally
convergent algorithms, are discussed in \cite{global_broyden}.

\subsection{Modified Broyden's Method}\label{MBM}

If the self-consistent equations involve many variables, i.e., $N$ is large, 
  storing the $N\times N$  matrix elements of the inverse 
Jacobian  and performing the $N\times N$ matrix multiplications
might be prohibitive; hence, further improvements are needed.
To deal with the size issues and to improve performance, Broyden's
method has been modified \cite{vanderbilt,mbm,Eyert}.  The
modification by Johnson \cite{mbm}
is now widely used in quantum chemistry. This {\it modified Broyden's
method} only utilizes information obtained in  $M$ previous iterations,
and this information is used in the update of the approximate inverse Jacobian.
 We recall here the final expressions of this modified Broyden
mixing procedure (a detailed derivation is given in Ref.~\cite{mbm}):
\begin{equation}
\begin{array}{c}
\Vvec_{\text{in}}^{(m+1)} \displaystyle = \Vvec_{\text{in}}^{(m)}+ \alpha \Fvec^{(m)}-
\sum_{n=\tilde{m}}^{m-1} w_n 
\gamma_{m n} \uvec^{(n)},
\end{array}
\label{BM}
\end{equation}
with $\tilde{m} = \max(1,m-M)$ and
\begin{equation}
\begin{array}{c}
\displaystyle \gamma_{m n}=\sum_{k=\tilde{m}}^{m-1} c_k^m \beta_{kn},~~~~
\beta_{kn}=(w_0^2 \mat{I} +\mat{a})^{-1}_{kn}, 
\medskip\\
\displaystyle c_k^m = w_k \left(\Delta\Fvec^{(k)}\right)^\dag \Fvec^{(m)},
\medskip\\
\displaystyle \mat{a}_{kn} =w_k w_n \Delta F^{(k)} \left(\Delta F^{(n)}\right)^\dag,
\end{array}
\label{BM1}
\end{equation}
where
\begin{equation}
\begin{array}{c}
\displaystyle \uvec^{(n)}=\alpha \Delta \Fvec^{(n)}+\Delta \Vvec^{(n)} \\~\\
\displaystyle 
\Delta \Vvec^{(n)}=\frac{\Vvec_{\text{in}}^{(n+1)}-\Vvec_{\text{in}}^{(n)}}{|\Fvec^{(n+1)}-\Fvec^{(n)}|}, 
\medskip\\
\displaystyle 
\Delta \Fvec^{(n)}=\frac{\Fvec^{(n+1)}-\Fvec^{(n)}}{|\Fvec^{(n+1)}-\Fvec^{(n)}|}.
\end{array}
\label{BM2}
\end{equation}
The weights $w_n$ ($n=1,...,M$) are associated with each previous
iteration and the values $w_n=1$ are usually chosen. The weight $w_0$
is assigned to the error in the inverse Jacobian and the value
$w_0=0.01$ proposed in \cite{mbm} gives stable results. The first two
terms in Eq.~(\ref{BM}) are simply the linear mixing of Eq.~(\ref{LM})
with a mixing parameter $\alpha$, while the last term is an additional
correction. As mentioned in \cite{mbm}, the parameter $\alpha$ can be
chosen to be rather large (in the examples of
this paper we have used $\alpha=0.7$) compared
to the usual values used in the simple mixing approach.

Equations (\ref{BM}-\ref{BM2}) are all that is required to update $\Vvec$. 
In the modified Broyden's method, large 
 $N\times N$ matrices do not appear. It is only
 one $M\times M$
matrix $\mat{a}_{kl}$ and $M$ vectors of length $N$ that need to be stored.
The iterative procedure 
starts with an initial input guess $\Vvec_{\text{in}}^{(0)}$ which
generates the first output solution $\Vvec_{\text{out}}^{(0)}$. In the
next iteration, $m$=1,  $\Vvec_{\text{in}}^{(1)}$ is a linear combination
of $\Vvec_{\text{in}}^{(0)}$ and $\Vvec_{\text{out}}^{(0)}$ and the
correction term is zero. For $m$=2,  the new $\Vvec_{\text{in}}^{(2)}$
contains the Broyden correction including the information obtained in  the
previous ($m$=1) iterations. The process is repeated until convergence
is reached.  

Our actual implementation of the modified Broyden's method is based on a
modified
subroutine  from the package \PWscf\
(Plane-Wave Self-Consistent Field)
  for electronic-structure \DFT\
calculations  using pseudo-potentials and a plane-wave basis set
\cite{pwscf}. The code is published under the \GNU\ General Public
License. The Broyden mixing routine in \PWscf\ 
is a simple module of about 20 statements
that uses the \BLAS\ and \LAPACK\ libraries.

\section{Spherical DFT Theory With Mean-Field Mixing: Two-Component  Unitary Fermi Gas}
\label{sec:unitary-fermi-gas}
As a first application of the standard Broyden's method in nuclear structure
calculations, we describe the numerical solution of a two-component
superfluid in a spherically symmetric harmonic trap using  the
superfluid local density approximation ({\SLDA}) of {\DFT}. The system is
described by a three-parameter local energy density functional \cite{unitaryGas}
adjusted to results
of Monte-Carlo calculations 
for homogeneous matter. The functional
depends on the three local densities:
$\rho(\boldsymbol{r})$ (particle density),  $\tau(\boldsymbol{r})$
(kinetic energy density),
and  $\kappa(\boldsymbol{r})$ (pairing tensor), which
 are constructed from the 
usual Bogoliubov
two-component quasi-particle wave functions $[U_{k}(\boldsymbol{r}), V_{k}(\boldsymbol{r})]$:
\begin{gather}
  \begin{aligned}
    \rho_{\uparrow}(\boldsymbol{r}) &= \sum_{k} |U_{k}|^{2}f_{T}(E_{k}), & 
    \rho_{\downarrow}(\boldsymbol{r}) &= \sum_{k} |V_{k}|^{2}f_{T}(-E_{k}),\\
    \tau_{\uparrow}(\boldsymbol{r}) &= \sum_{k} |\nabla U_{k}|^{2}f_{T}(E_{k}), & 
    \tau_{\downarrow}(\boldsymbol{r}) &= \sum_{k} |\nabla V_{k}|^{2}f_{T}(-E_{k}),
  \end{aligned}\nonumber\\
  \label{eq:anomalous_density}
  \kappa(\boldsymbol{r}) =  \sum_{k}U_{k}V_{k}^{*}\frac{f_{T}(-E_{k})-f_{T}(E_{k})}{2}
\end{gather}
where 
$k$ runs over the
quasi-particle states, $\rho=\rho_{\uparrow}+\rho_{\downarrow}$,
 and the function $f_{T}(E) =
[1+\exp(E/T)]^{-1}$ is the thermal distribution function for
fermions. 
The resulting Kohn-Sham equations
can be written in the usual \HFB\ form,
\begin{equation}\label{hfbm}
  \mathcal{H}\left[h(\boldsymbol{r}),\Delta(\boldsymbol{r})\right]
  \begin{pmatrix}
    V_{k}(\boldsymbol{r})  \\
    U_{k}(\boldsymbol{r}) 
  \end{pmatrix}
  =
  E_{k}
  \begin{pmatrix}
    V_{k}(\boldsymbol{r})  \\
    U_{k}(\boldsymbol{r}) 
  \end{pmatrix},
\end{equation}
with the {\HFB} matrix given by
\begin{equation}\label{matrixh}
  \mathcal{H} = 
  \begin{pmatrix}
    h(\boldsymbol{r}) - \mu_{\uparrow} 
    & \Delta(\boldsymbol{r})\medskip\\
    \Delta^{\dagger}(\boldsymbol{r}) 
    & -h(\boldsymbol{r}) + \mu_{\downarrow}
  \end{pmatrix},
\end{equation}
where the particle-hole potential $\Gamma(\boldsymbol{r})$, which
enters the particle-hole Hamiltonian
\begin{equation}\label{effmass}
h(\boldsymbol{r}) = -\tfrac{\hbar^2}{2m^*}\nabla^{2} + \Gamma(\boldsymbol{r}),
\end{equation}
and the pairing potential  $ \Delta(\boldsymbol{r})$ are both
functions of the local densities. 
(The value of the effective mass $m^*$ in Eq.~(\ref{effmass}) has been 
determined in Ref.~\cite{unitaryGas}.)
The introduction of two chemical potentials $\mu_{\uparrow}$ and
$\mu_{\downarrow}$ eschews the need to use a different formalism
(Pauli blocking) for dealing with odd systems \cite{unitaryGas,Randeria}.

The spherical
 \HFB\ equations (\ref{hfbm}) 
 are solved by discretizing $\mathcal{H}$ 
in the radial coordinate $r$ using  
the so-called \DVR\ basis \cite{DVR-1,DVR-2,DVR-3}.  We use two sets
of abscissa -- $\{r_{0,n}\}$ ($n=1,\ldots, N_r$) for even-$\ell$ 
 and $\{r_{1,n}\}$ for odd-$\ell$ partial waves.
Thus, to represent the functions $\Delta(r)$,
$\Gamma(r)$, etc., we need only store $2N_r$ sets of data.

The chemical
potentials for the two species ($\mu_{\uparrow}$ and
$\mu_{\downarrow}$) can be viewed as the  Lagrange multipliers that
enter  the grand
thermodynamic potential $\mathcal{E} -
\mu_{\uparrow}\rho_{\uparrow} -
\mu_{\downarrow}\rho_{\downarrow}$. 
In the asymmetric situation, $\delta\mu=\mu_{\uparrow}-\mu_{\downarrow}\ne 0$,
 the spectra for
the species are shifted so that the thermal distribution functions
$f_{T}(E)$ appearing in~(\ref{eq:anomalous_density})  results in
different numbers of each species.  At zero temperature this produces 
a quantized step as $\delta\mu$ approaches the pairing gap. By
introducing a small temperature, however, this discontinuity can be smoothed
without noticeably affecting the physics.
This smoothing is important in order for Broyden's method 
to achieve  convergence.

Instead of treating the chemical potentials as external parameters and
adjusting them separately during the iteration process, we have found that
they may be adjusted \emph{during the iteration} in
such a way that the process converges to a solution with the desired
particle number constraints. 
After some experimenting, we have found that -- in order to speed up the convergence --
the chemical potentials should be updated after every iteration according to a 
simple ansatz:
\begin{equation}
  \label{eq:mu_update}
  \mu_{\sigma} \mapsto \mu_{\sigma} + 
  \eta \frac{N^{0}_{\sigma}-N_{\sigma}}{N^{0}_{+}}\mu_{TF}(N^{0}_{\sigma}),
\end{equation}
where $\sigma=\downarrow$ or $\uparrow$, $N^{0}_{\sigma}$ is the desired particle number,
and $\mu_{TF}(N)$ is some typical positive scale for
the chemical potentials, and $N^{0}_{+}$ is some typical positive scale
for the particle numbers (such as the total desired particle number). The parameter $\eta$ can be adjusted to ensure convergence.
 Note
that the computed particle number $N_{\sigma}$ enters in exactly one
place so that the update is zero if and only if $N_{\sigma} =
N^{0}_{\sigma}$.  Thermodynamics ensures that the chemical potential is
adjusted in the appropriate direction, so as long as the scales
in Eq.~(\ref{eq:mu_update}) are appropriately chosen the process
converges nicely.  The only problems we have encountered are: (i) If the
updated steps  are too large, the system might oscillate wildly.  This is easily
compensated for by choosing a smaller value of $\eta$; (ii) If the
temperature is too low, the resulting function may be discontinuous,
in which case the update (\ref{eq:mu_update}) may cause an oscillation.
Introducing a small, but finite, temperature (smaller than any relevant
scale)  cures this problem.

The Broyden vector $\Vvec^{(m)}$ in our {\SLDA} calculations contains
the values of $\Gamma(r)$ and $\Delta(r)$  
 on two \DVR\ grids, and  two chemical potentials:
\begin{equation}
\Vvec \equiv \left\{ \Gamma(r_{0,n}), \Gamma(r_{1,n}),
  \Delta(r_{0,n}), \Delta(r_{1,n}), \mu_{\downarrow},
  \mu_{\uparrow}
\right \}.
\end{equation}
 Thus,
for a basis of size $N_r$, the size of the Broyden vector is $N=4N_r + 2$.

We found that in order to ensure consistency in the above  
scheme, \emph{all} parameters used in the
iteration should appear in the Broyden process. Thus, even though our
description in terms of the \DVR\ basis is somewhat redundant with the
function evaluated at two sets of abscissa $\{r_{0,n}\}$ and
$\{r_{1,n}\}$, one must keep track of all the information describing
the  Jacobian. Also, one must be extremely
careful to include any external parameters that are adjusted during
the iteration in the Jacobian, as well as any internal parameters that
are allowed to relax, but which might add hysteresis to the
system. The output of a single iteration must be a smooth and
uniquely defined function of only the input vector
$\Vvec^{(m)}_{\rm in}$.
\begin{figure}[t]
  \includegraphics[width=\columnwidth]{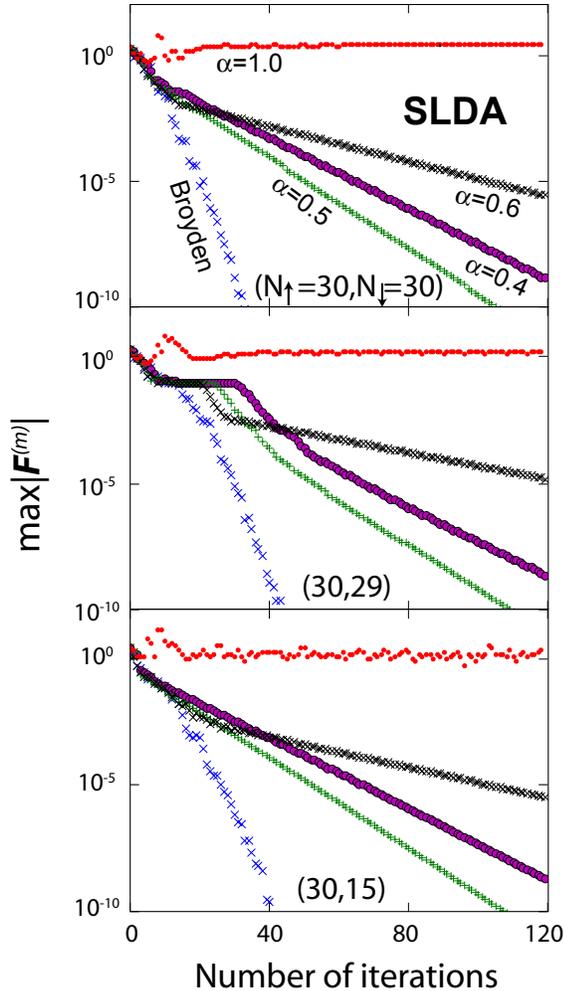}
  \caption{\label{fig:SLDA} 
  (Color online)
  Comparison between the 
  linear mixing algorithm   and the standard Broyden's method
  in {\SLDA} for systems of
    $(N_{\uparrow},N_{\downarrow})=(30,30)$ (top),
    $(N_{\uparrow},N_{\downarrow})=(30,29)$ (middle), and
    $(N_{\uparrow},N_{\downarrow})=(30,15)$ (bottom).  The basis size
    is $N_r$=40 and the convergence parameter
    of Eq.~(\ref{eq:mu_update}) is $\eta$=1. 
    For the asymmetric systems, the equations exhibit a
    discontinuity responsible for  small plateaus in the
    plots. Introducing a small temperature $T=0.02\epsilon_{F}$
    restores the continuity.  Once the approximate solution is found,
    the temperature may be set back to zero.}
\end{figure}

Figure \ref{fig:SLDA} displays some typical results. Here we solve
the problem of  60, 59, and 45 particles in a harmonic trap. We consider the
cases with equal numbers $(N_{\uparrow}, N_{\downarrow}) = (30, 30)$, one particle extra
$(N_{\uparrow}, N_{\downarrow}) = (30, 29)$, and 
half-paired $(N_{\uparrow}, N_{\downarrow}) = (30, 15)$.
The size of the Broyden vector is $N$=162.
We see that in all cases, Broyden's method performs substantially better
than the linear mixing  method.

Note that there are many cases in which the full-step 
algorithm ($\alpha$=1)
does not
converge. This is typically because the system exhibits some form of
oscillation and the full step overshoots each time. In our problem
this is due to the chemical potential update, and indicates that we
should probably choose a smaller factor $\eta$ in
Eq.~(\ref{eq:mu_update}).  Broyden's method, however,
automatically detects this and compensates accordingly,
thus achieving good convergence even if $\eta$ is not properly chosen.

In the odd-particle  case (30, 29),  
a plateau appears where the convergence improves very slowly. This
marks the point at which the chemical potentials adjust so as to split
one state off from the Fermi sea. Prior to this, the large-scale
structure of the initial state is corrected, but once this occurs, the
chemical potentials must be finely tuned so that exactly one extra
state is occupied. Broyden's method has the same problem because 
 the approximate gradient changes rapidly.
To deal with this,  a small temperature is assumed. This makes the
functions smooth and facilitates convergence. Once the steps are small
enough that the energies do not cross zero, the oscillations terminate and
convergence is resumed. If problems persist, the
strategy is to start with a fairly large temperature to isolate the
approximate values of the chemical potentials (only achieving rough
convergence), and then iterating with these values at much lower
temperatures to reach the desired tolerance.

\section{Symmetry-unconstrained HFB Theory With Mean-Field Mixing}
\label{sec:symm-unconstr-hfb}

In this section
we discuss the adaptation and performance of the modified Broyden's method in
self-consistent nuclear structure calculations
 with the \HFODD\ code of Ref.~\cite{hfodd,hfodd-1}. The code
solves the Skyrme-\HFB\ problem by
expanding the solutions on the anisotropic  cartesian harmonic
oscillator (\HO) basis. {\sc hfodd} does not have any
self-consistent symmetry built in; hence, all deformation degrees of
freedom can be taken into account. As the time-reversal symmetry can be
broken in {\HFODD}, the code can be used to describe
 nuclear rotation using the
cranking approximation. 

The building blocks of the  Skyrme-\HFB\ approach
are the particle density 
$\rho(\boldsymbol{r},\boldsymbol{r}')$ 
and the pairing density  $\tilde{\rho}(\boldsymbol{r},\boldsymbol{r}')$.
The latter 
is related to the  pairing tensor 
$\kappa$ of Sec.~\ref{sec:unitary-fermi-gas}
by an anti-unitary transformation  (see, e.g., \cite{Dob96}).

From 
$\rho$ 
and  $\tilde{\rho}$,
other  local
densities and currents  are constructed by considering spin and isospin degrees of
freedom, and by taking derivatives up to the second order \cite{LDA}. 
By coupling
the local  densities and currents, one constructs  the  energy density
$\mathcal{E}(\boldsymbol{r})$ that is scalar,
time-even, and isoscalar.
In its current version, {\sc hfodd} does
not consider explicit proton-neutron mixing; hence, the \HFB\ equations are solved
for protons and neutrons separately. 
A variation of the total energy with respect to the single-particle
wave functions yields the mean field $h$ and pairing field $\tilde{h}$
that enter the \HFB\ Hamiltonian matrix: 
\begin{equation}
\mathcal{H} = \left(
\begin{array}{cc}
h(\boldsymbol{r})-\lambda & \tilde{h}(\boldsymbol{r}) \\
\tilde{h}(\boldsymbol{r}) & -h(\boldsymbol{r})+\lambda
\end{array}
\right).
\label{HFBMatrix}
\end{equation}
 The particle-hole
mean-field Hamiltonian can be expressed as:
\begin{equation}
h(\boldsymbol{r}) = -\frac{\hbar^{2}}{2m}\Delta + \Gamma_{0}^{even} + \Gamma_{0}^{odd} 
+ \Gamma_{1}^{even} + \Gamma_{1}^{odd} 
+ U^{Coul},
\label{mean-field}
\end{equation}
where the subscript $t$=0 (1) denotes isoscalar (isovector) fields, while the superscript 
$even$ ($odd$) labels time-even (time-odd) fields.

The pairing field
$\tilde{h}$ in {\sc hfodd} is related to the usual pairing field
$\Delta$ in Eq.~(\ref{matrixh}) by the same anti-unitary transformation 
that relates $\tilde{\rho}$ and $\kappa$. In our \HFB\ calculations,
we employed the density-dependent delta interaction in the particle-particle channel.
The corresponding particle-particle mean-field Hamiltonian reads:
\begin{equation}
\tilde{h}(\boldsymbol{r}) = \frac{1}{2}V_0\left[1-V_1\frac{\rho(\boldsymbol{r})}{\rho_0}\right] 
\tilde{\rho}(\boldsymbol{r}),
\label{pairing-field}
\end{equation}
where 
$\rho_0$=0.16\,fm$^{-3}$. It this paper we adopted the
 value of $V_1$=0.5 corresponding
to the so-called mixed pairing  (see Ref.~\cite{Dob02} and
references therein).
Due to the zero-range of the pairing interaction, a cut-off energy
of 60\,MeV is used when summing up the contributions of the \HFB\
quasi-particle states to the density matrices. For a given cut-off,
the pairing strength $V_{0}$ is determined such as
to reproduce the  experimental neutron pairing gap in $^{120}$Sn,
$\Delta_{n}$=1.245\,MeV. As discussed in Ref.~\cite{Bor06}, this renormalization
procedure gives very similar results as  the pairing regularization
method \cite{Bul02}.

The mean-field Hamiltonian (\ref{mean-field}) depends on the following self-consistent
potentials built of the local densities and currents:
\begin{equation}
\begin{array}{ll}
\displaystyle
\Gamma_{t}^{even}(\boldsymbol{r}) & = \displaystyle
-\boldsymbol{\nabla}\cdot\left[ M_{t}(\boldsymbol{r})\boldsymbol{\nabla}\right]
+ U_{t}(\boldsymbol{r}) \medskip\\
                  & \displaystyle
+ \frac{1}{2i} \left( 
\overleftrightarrow{\nabla}\sigma\cdot\overleftrightarrow{B}_{t}(\boldsymbol{r}) + 
\overleftrightarrow{B}_{t}(\boldsymbol{r})\cdot\overleftrightarrow{\nabla}\sigma
\right) \medskip \\
\Gamma_{t}^{odd}(\boldsymbol{r}) & = \displaystyle
-\boldsymbol{\nabla}\cdot\left[ \left( \boldsymbol{\sigma}\cdot\boldsymbol{C}(\boldsymbol{r})\boldsymbol{\nabla}\right)\right]
+ \boldsymbol{\sigma}\cdot\boldsymbol{\Sigma}_{t}(\boldsymbol{r}) \medskip\\
                  & \displaystyle
+ \frac{1}{2i} \left( 
\boldsymbol{\nabla}\sigma\cdot\boldsymbol{I}_{t}(\boldsymbol{r}) + 
\boldsymbol{I}_{t}(\boldsymbol{r})\cdot\boldsymbol{\nabla}
\right).
\end{array}
\end{equation}
The potentials $U_{t}(\boldsymbol{r})$ and $M_{t}
\boldsymbol{r})$ are real scalar fields (1 component),
$\boldsymbol{\Sigma}_{t}(\boldsymbol{r})$,
$\boldsymbol{C}_{t}(\boldsymbol{r})$ and
$\boldsymbol{I}_{t}(\boldsymbol{r})$ are real vector fields (3
components), and $\overleftrightarrow{B}_{t}(\boldsymbol{r})$ is a real
second-rank tensor field (9 components). The pairing potential 
$\tilde{h}_{t}(\boldsymbol{r})$ is a two-component complex scalar field.

The \HO\ eigenfunctions are Hermite polynomials
weighted by a Gaussian factor. Consequently, all
integrations in  {\sc hfodd} are  carried out by using the Gauss-Hermite quadrature
\cite{hfodd}. The mesh of the Gauss-Hermite nodes defines the
discretized grid, and it is sufficient to define  mean-field
potentials on this grid to compute the
\HFB\ matrix. 

The Broyden vector $\Vvec^{(m)}$ in \HFODD\ contains the values of the 44
self-consistent field components on the Gauss-Hermite grid
$\boldsymbol{r_i}$:
\begin{multline}
\Vvec \equiv \left\{U_{t}(\boldsymbol{r_i}), M_{t}(\boldsymbol{r_i}),
\boldsymbol{\Sigma}_{t}(\boldsymbol{r_i}), \boldsymbol{C}_{t}(\boldsymbol{r_i}),
\boldsymbol{I}_{t}(\boldsymbol{r_i}),\right.\\
\left. \overleftrightarrow{B}_{t}(\boldsymbol{r_i}), \Re[\tilde{h}_{t}(\boldsymbol{r_i})], 
\Im[\tilde{h}_{t}(\boldsymbol{r_i})]
   \right \}.
\end{multline}
The
advantage of dealing with the fields is that one does not have to
worry about the unitarity condition for the generalized density matrix.

If $N_{x}$, $N_{y}$, and $N_{z}$ are the numbers of Gauss-Hermite nodes in the $x$-, 
$y$- and $z$-direction respectively, the total size of the Broyden vector in \HFODD\
is $N = (N_{x}N_{y}N_{z})\times 44$.
For a spherical \HO\ basis with $N_{osc}$=12, the number of Gauss-Hermite
points required to perform exact integrations is $N_{x} = N_{y} = N_{z}
= 26$, and the size of the Broyden vector becomes $N$=773,344. In
double precision floating point arithmetics, this represents
approximately 6.2\,MB of memory. For  $M$=8  vectors 
retained in the Broyden history,
 this represents a  100\,MB memory overhead.
Calculations involving larger bases are often required for
studies of, e.g., fission pathways, and it is not unusual to deal with
$N_{osc}$=20 for which the size of the Broyden vector
becomes as large as  $N$=3,259,872, corresponding to  
$\sim$416\,MB of additional memory.

\begin{figure}[htb]
\includegraphics[width=0.47\textwidth]{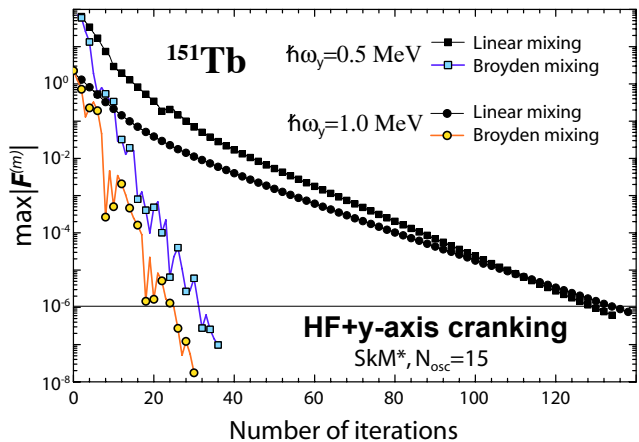}
\caption{(Color online) 
Comparison between linear mixing ($\alpha$=0.5, black symbols)  and the
modified Broyden's method
($\alpha$=0.7, $M$=7) in \HFODD \ for  a high-spin superdeformed band in  $^{151}$Tb at two 
values of  rotational 
frequencies $\hbar\omega_y$=0.5\,MeV and 1\,MeV. 
	    }
\label{fig2}	
\end{figure}

Figure~\ref{fig2} illustrates the use of Broyden's method for a
cranked Hartree-Fock calculation (without pairing)
for the lowest high-spin  superdeformed band of $^{151}$Tb.
The \textsc{skm}$^{*}$ Skyrme functional \cite{SkM} has been used. The rotation is generated by adding to
the mean-field
Hamiltonian (\ref{mean-field})  the cranking term
$-\omega_{y}\hat{J}_{y}$, with $\omega_y$ being the  rotational frequency.
As the time-reversal symmetry is broken in {\HF}, both time-even and time-odd
fields appear.
The calculation at $\hbar\omega_{y}$=0.5\,MeV
was initialized with the unconstrained \HF\ results for the
superdeformed vacuum configuration of $^{151}$Tb (with all the lowest
single-particle \HF\ routhians
occupied) at $\omega_{y}$=0. The calculations at $\hbar\omega_{y}$=1.0\,MeV
were
initialized with the converged results at  $\hbar\omega_{y}$=0.5\,MeV. 
The same  stretched \HO\ basis containing  $N_{osc}$=15
shells and a deformation of $\beta_{2}$=0.61 and $\beta_{4}$=0.1 
was used in both cases.
The advantage of the modified
Broyden's method over the simple mixing is evident. While in the standard
calculations the required tolerance of $10^{-6}$ is achieved after $m$$\sim$30 iterations, 
using the linear  mixing, the same outcome is achieved after 130 steps. 

\begin{figure}[htb]
\includegraphics[width=0.47\textwidth]{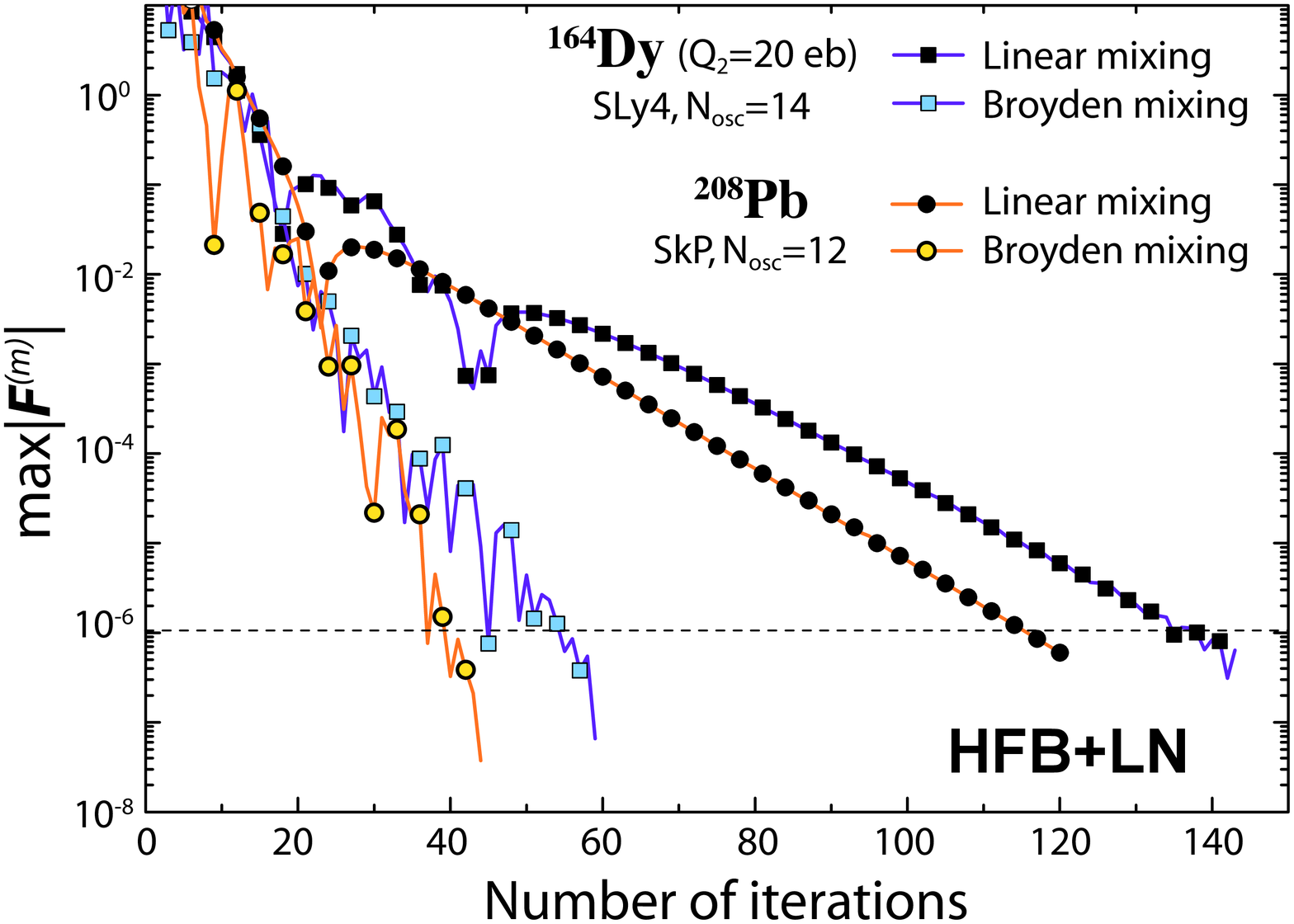}
\caption{(Color online) 
Comparison between linear mixing ($\alpha$=0.5, black symbols)  and Broyden's method
($\alpha$=0.7, $M$=7)
in \HFODD\ 
 for the superdeformed minimum in
$^{152}$Dy at the quadrupole moment of $Q_{2}$=20\,eb (SLy4 Skyrme functional, $N_{osc}$=14)
and the spherical ground state of 
$^{208}$Pb (SkP Skyrme functional, $N_{osc}$=12). 
	    }
\label{fig3}	
\end{figure}

Figure~\ref{fig3} illustrates the 
 constrained \HFODD\ calculations  with pairing for the superdeformed minimum in
$^{152}$Dy at the quadrupole moment of $Q_{2}$=20\,eb
(using the SLy4 Skyrme functional \cite{SkL})
and also for  the spherical ground-state of $^{208}$Pb
(using the SkP Skyrme functional \cite{SkP}). Both
calculations were carried out with mixed pairing and the Lipkin-Nogami (\LN) particle-number
corrections (see, e.g., Ref.~\cite{Sto07,dftresults}).
In order to obtain stability of the results, the proton and neutron $\lambda_{2}$ parameters 
of the \LN\ model had to be incorporated into the Broyden vector. 
Again, as in Fig.~\ref{fig2}, the modified Broyden's method 
provides an impressive improvement over the linear mixing scheme.

Before concluding this section, we would like to make an important
technical remark. In cartesian coordinates, the \HO\ wave-functions are
products of the Hermite polynomials and the Gaussians. In the code
\HFODD\ the Gaussians do not have to be calculated explicitly, as they
are already incorporated into the Gauss-Hermite quadrature weights
\cite{hfodd}. However, this implies that the mean fields used in \HFODD\
are Gaussian-scaled. Consequently, they can take non-zero (and possibly
large) values at some points of the Gauss-Hermite grid  where the
physical potentials practically vanish. Using the scaled potentials as
inputs to Broyden's method can be a source of very significant
numerical instability. Indeed if the scaled potentials (or densities)
are used, numerical and physically insignificant fluctuations of the
real potentials are artificially enhanced, thereby introducing
considerable (and artificial) numerical  noise. This instability is not
present  when the physical mean-field potentials are used.

\section{Symmetry-constrained HFB Theory with Hamiltonian Matrix Mixing}
\label{sec:symm-constr-hfb}

In this section we discuss the performance  of the modified Broyden's
method as applied to the axial  \HFB\  problem.
To this end, we employ the \HFBTHO\ code \cite{dftresults,hfbtho} that solves the
Skyrme-\HFB\ equations by expanding the eigenstates in the cylindrical
transformed harmonic oscillator  (\THO) basis. In the examples  shown in
this section, we use a SLy4 Skyrme energy density functional augmented
by mixed  pairing (\ref{pairing-field}). The 
 \HFB\ Hamiltonian matrix (\ref{HFBMatrix}) is
represented in a \THO\  basis. 

The Broyden vector $\Vvec^{(m)}$ in \HFBTHO\ 
contains the matrix elements of the
self-consistent fields $h$ and $\tilde{h}$  for neutrons and protons:
\begin{equation}\label{mfor}
\Vvec \equiv \left\{ h^n_{ij}, h^p_{ij}, \tilde{h}^n_{ij}, \tilde{h}^p_{ij} \right \}.
\end{equation}
Since $h$ and $\tilde{h}$ are hermitian, $i \le j$.
The
chemical potentials for neutrons and protons are adjusted at each step
$m$ to the correct number of particles. 
The matrix formulation (\ref{mfor}) makes it easy to
efficiently incorporate non-local terms that appear,
e.g., in the  particle-number restoration schemes.
Some  results of
our Broyden \HFBTHO\  calculations have been reported in Ref.~\cite{Stob}.

Since \HFBTHO\  strictly preserves axial symmetry, parity, and time
reversal,  the number of
independent matrix elements is significantly reduced as compared to
the full symmetry-unconstrained case of Sec.~\ref{sec:symm-unconstr-hfb}.
For instance, for
the basis space of $N_{\rm osc}$=20 principal oscillator shells, the
length of the Broyden vector is $N$=261,228. 
For the modified
Broyden's method with  $M$=8, the required memory increase in  \HFBTHO\
as compared to  the simple 
linear mixing algorithm
is usually less than
33\,MB, i.e., is practically negligible.

In the implementation of the  linear mixing procedure in {\HFBTHO}, the
mixing  parameter $\alpha$ varies. Initially its value is set at
$\alpha$=0.1. In the next iteration, if the  largest element of
$\Fvec^{(m)}$,  max$|\Fvec^{(m)}|$, is less than  max$|\Fvec^{(m-1)}|$,
$\alpha$ is multiplied by a factor of 1.13 until it reaches the maximum
value of $\alpha$=1.0. When the condition
max$|\Fvec^{(m)}|$$>$max$|\Fvec^{(m-1)}|$ is met, $\alpha$  is returned to its
initial value of 0.1, and the process starts all over again. Such a
technique has yielded stable results for all cases considered.

\begin{figure}[htb]\centering
\includegraphics[width=0.47\textwidth]{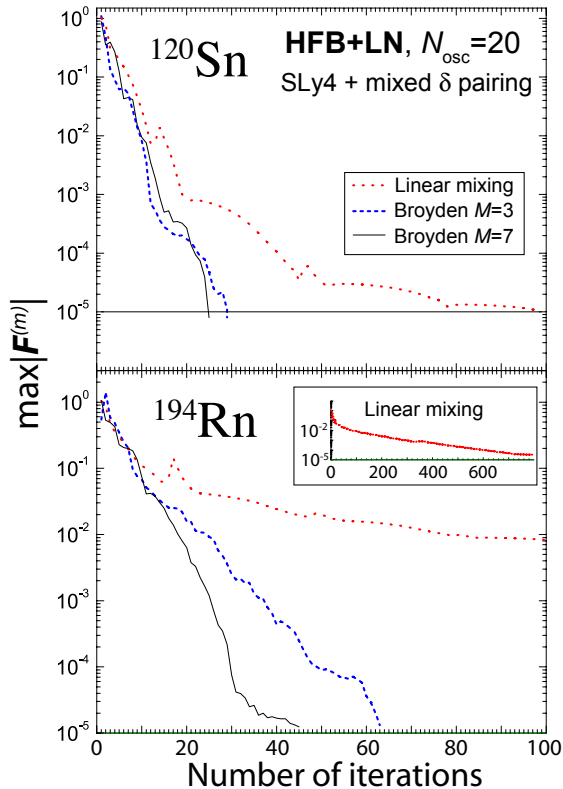}
\caption{(Color online) 
Comparison between linear mixing (dotted line)  and Broyden's method
in \HFBTHO\ for  $^{120}$Sn (top) and  $^{194}$Rn (bottom). The largest
element of $\Fvec^{(m)}$ is shown  as a function of $m$. The modified
Broyden mixing  was carried out with $\alpha$=0.7, $w_0$=0.01, and
$w_n$=1 for  $M$=3 (dashed line) and $M$=7 (solid line). The inset shows
the slow convergence of the linear mixing for the pathological case of
$^{194}$Rn. Here, the
tolerance 10$^{-5}$ is reached  after 4,345 iterations.
}
\label{fig4}
\end{figure}

Figure \ref{fig4} compares linear mixing  and Broyden's method
for the  spherical nucleus $^{120}$Sn, which is often
used to determine the pairing strength of the functional. Both
calculations are initiated from the same  Woods-Saxon fields
\cite{hfbtho}. 
Usually convergence of all physical quantities of interest is achieved when 
max$|\Fvec^{(m)}|$ is less than $10^{-5}$.
For the nucleus $^{120}$Sn, which represents a typical case, 
Broyden's method increases the  convergence rate by a factor of
$\sim$3,
as compared to the linear mixing. While the absolute convergence
rate does depend on an individual nuclear configuration,
Broyden's method has proved faster than the linear mixing in all the
cases investigated. In particular, for some pathological configurations the
improvement is staggering. Such an example is shown in Fig.~\ref{fig4}
(bottom) for the prolate configuration in $^{194}$Rn.
Here, the efficiency of Broyden's algorithm
is almost a hundred times better than the linear mixing. An
additional 30\% gain can be made by increasing the number of vectors
retained in the Broyden history from $M$=3 to $M$=7.

\begin{figure}[htb]\centering
\includegraphics[width=0.47\textwidth]{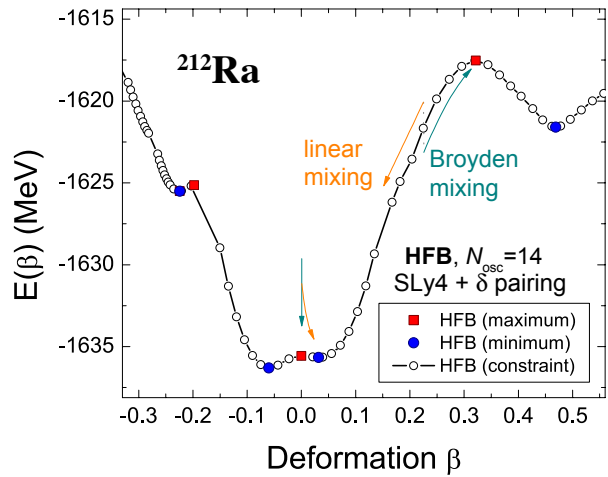}
\caption{(Color online) 
Energy curve of 
$^{212}$Ra versus quadrupole deformation 
obtained in the constrained \HFBTHO\ calculations. 
The local minima (maxima) are marked by dots (squares). While in the
unconstrained calculations
the  linear mixing converges at the local minima only, Broyden's method yields {\it both}
local minima and maxima, as indicated by arrows.
}
\label{fig6}
\end{figure}

Broyden's  method proves especially helpful for constrained
calculations in which the total energy is calculated as a function of
some collective variables (usually expectation values of selected
one-body operators).  A typical example is shown in 
Fig.~\ref{fig6} that displays the quadrupole deformation energy curve for
$^{212}$Ra. The quadrupole deformation $\beta$ is related to the total quadrupole
moment $Q_{20}$ via the usual relation:
\begin{equation}
Q_{20} = \sqrt{5 \over \pi} \langle r^2 \rangle \beta.
\end{equation}
 Starting from the converged solution at the
neighboring  point of the deformation curve, the calculation converges 
in no more than 20-30 iterations. The number of iterations is
reduced by a factor of 2-3 with respect to the linear mixing used under the
same conditions. Since constrained calculations involve a
large number of  points along the collective path, 
and  in many applications several collective
coordinates are involved, such a gain greatly
improves the scalability of computations.

A feature of Broyden's method, which is not present when using the 
linear mixing, is the ability to converge at both minima and maxima of the 
energy curve, which represent solutions of unconstrained \HFB\ calculations.
The extrema in the energy curve of $^{212}$Ra are marked in 
Fig.~\ref{fig6} by black symbols. 
Starting the unconstrained calculation 
from a  constrained solution at  $\beta$=0.25 and using
the linear mixing, one obtains  the local minimum at
$\beta$$\approx$0.05. Under the same conditions, 
the Broyden mixing yields the maximum at $\beta$$\approx$0.3.
A similar situation happens when starting with an initial guess at zero
deformation: the linear mixing converges to  the slightly deformed 
prolate minimum, while the Broyden mixing yields the  spherical maximum of
the energy curve.

This interesting  feature of Broyden's method may be particularly
useful for many-body tunneling  calculations (e.g., fission), 
since it allows to compute the  potential
barrier height with a much better accuracy than when using the linear
mixing in a constrained mode. 
However, it also implies that a more detailed knowledge of the
potential energy surface is required to compute unconstrained minima,
since some of the converged solutions may turn out to be local 
maxima. A possible
solution to this problem is to make use of the curvature condition
\begin{equation}
\begin{array}{c}
\displaystyle C^{(m)} =  \left(\Vvec_{\text{out}}^{(m-1)}-\Vvec^{(m)}\right)^\dag
\mat{B}^{(m)} \Fvec^{(m)} > 0
\end{array}
\label{CC}
\end{equation}
which guarantees that  the Broyden step is downhill, i.e., 
toward the  minimum. In practical calculations, we accept the Broyden
step if Eq.~(\ref{CC}) is satisfied. 
Otherwise, we apply the linear mixing while keeping the current
iteration in the Broyden history. Such a procedure bypasses 
maxima or inflection points and
properly converges to the nearest local minimum.

\section{Coupled-Cluster Applications}
\label{sec:coupl-clust-theory}

In this section we illustrate the efficiency of the modified Broyden's method
in the context of the nuclear \CC\ theory.
Coupled-Cluster theory originated in nuclear structure; it was
pioneered by Coester and K\"ummel \cite{Coe58,Coe60} in the early
1960s. In the last two decades, it has been applied mostly in quantum
chemistry and today it defines  the state-of-the-art many-body theory in
this field (see \cite{Bar07} for a recent overview). Recently, the \CC\
approach  has seen a revival in nuclear structure \cite{Dea05}. 
In \cite{Hag06} it was
applied to the description of loosely bound and unbound helium
isotopes; in \cite{Hag07a} it was shown that coupled cluster meets
few-body benchmarks, and converged results for the ground state energy
of $^{40}$Ca were given; in \cite{Hag07b} coupled-cluster theory was
extended to three-body Hamiltonians, and the first \CC\ calculations
with three-nucleon forces were presented.

Within the \CC\ theory, the nuclear ground-state wave function is written as
\begin{equation}
\label{cc_ansatz}
|\psi\rangle = e^{\hat{T}} |\phi\rangle \,,
\end{equation}
where $|\phi\rangle=\prod_{i=1}^A \hat{a}_i^\dagger|0\rangle$ is a
single-particle product state expressed in some convenient
representation and
\begin{equation}
\label{T}
\hat{T} = \hat{T}_1 + \hat{T}_2 + \ldots + \hat{T}_A
\end{equation}
is a correlation operator that can be expressed in terms of $k$-body operators 
\begin{equation}\label{Top}
\hat{T}_k =
\frac{1}{(k!)^2} \sum_{i_1,\ldots,i_k; a_1,\ldots,a_k} t_{i_1\ldots
i_k}^{a_1\ldots a_k}
\hat{a}^\dagger_{a_1}\ldots\hat{a}^\dagger_{a_k}
\hat{a}_{i_k}\ldots\hat{a}_{i_1} 
\end{equation}
representing particle-hole (p-h) excitations with respect to the reference state
$|\phi\rangle$.
In Eq.~(\ref{T}) and in the following, $i, j, k,\ldots$ label hole
orbitals, while $a, b, c,\ldots$ refer to particle
states.

The many-body correlations atop  the  reference state are
introduced through the exponentiated correlation operator.
If no
truncations are introduced in $\hat{T}$, the theory is exact
and the choice of starting
reference state is arbitrary. However, in practice, one typically
restricts expansion (\ref{T})  to the 
two leading terms,
 $\hat{T} \approx  \hat{T}_1+\hat{T}_2$. 
Within the resulting coupled-cluster singles and doubles (\CCSD) model,
the  \CC\ equations can be written as
\begin{eqnarray}
\label{ccsd}
E &=& \langle \phi | \overline{H} | \phi\rangle\, , \medskip\\
\label{ccsd1}
0 &=& \langle \phi_i^a | \overline{H} | \phi\rangle\, , \medskip\\
\label{ccsd2}
0 &=& \langle \phi_{ij}^{ab} | \overline{H} | \phi\rangle \,.
\end{eqnarray}
Here $|\phi_{i_1\ldots i_n}^{a_1\ldots a_n}\rangle = \hat{a}_{a_n}^\dagger\ldots \hat{a}_{a_1}^\dagger \hat{a}_{i_1}\ldots \hat{a}_{i_n}|\phi\rangle$ is a $np-nh$ excitation of the reference state $|\phi\rangle$, and 
\begin{equation}
\label{hsim}
\overline{H} = e^{-\hat{T}} \hat{H} e^{\hat{T}} = \bigl( \hat{H} e^{\hat{T}} \bigr)_c
\end{equation}
is the similarity-transformed Hamiltonian (note that $\overline{H}$ is
non-Hermitian). The last expression on the right-hand side of
Eq.~(\ref{hsim}) indicates that only fully connected diagrams contribute
to the construction. This ensures that no unlinked diagrams enter the
\CC\ wave function, regardless of the truncation level in
$\hat{T}$, and this property makes \CC\ theory size
extensive \cite{Bar07}.

In order to determine the ground state energy $E$ in Eq.~(\ref{ccsd}), the
p-h excitation amplitudes $t_i^a$ and $t_{ij}^{ab}$ have to be
evaluated through a nonlinear
set of equations (\ref{ccsd1}) and (\ref{ccsd2}). By making use of
diagram rules and defining intermediates by writing the diagrams in
factorized form, the equations for the amplitudes can be written in a
quasi-linearized form. As an example, Eq.~(\ref{ccsd1}) can be written as:
\begin{multline}
  0 = f_a^i + I^{'e}_a t^i_e - \bar{h}^i_m t^m_a -  v^{ie}_{ma}t^m_e + 
\bar{h}^e_m t^{mi}_{ea} - {1\over 2}\bar{h}^{ie}_{mn}t^{mn}_{ae} \\
+ {1\over 2}v^{ef}_{am}t^{im}_{ef}. 
\label{eq:int_eq}
\end{multline} 
In (\ref{eq:int_eq}), $f_a^i$ is the Fock matrix and 
$I^{'e}_a$, $\bar{h}^i_m$, and $\bar{h}^{ie}_{mn}$ are the
intermediates \cite{gour06}.
The non-linearity is hidden in the  intermediates which depend 
on the amplitudes $t_i^a$ and $t_{ij}^{ab}$.

Typically the number of unknowns is too large to allow for a direct
solution of the \CC\ equations; hence, one
has to resort to iterative approaches. In the largest calculation we
have performed to date \cite{Hag07b}, we solved Eqs.~(\ref{ccsd1}) and
(\ref{ccsd2}) for $\sim 10^8$ number of unknowns, on 1000 2GB parallel
processors. Each iterative step costs, at the \CCSD\ level, $\sim 
n_h^2n_p^4$ computational cycles ($n_h/n_p$ is the number of hole/particle 
 orbitals). 
 
 Considering the number of unknowns and the highly non-linear nature of \CC\ equations,
the use of  iterative
convergence accelerators, such as the modified Broyden's method, is a must.
In our nuclear \CC\ calculations, the Broyden vector is given by the amplitudes:
\begin{equation}
\Vvec \equiv \left\{t_i^a, t_{ij}^{ab} \right\}.
\end{equation}

\begin{figure}[t]
\includegraphics[width=0.45\textwidth,clip=]{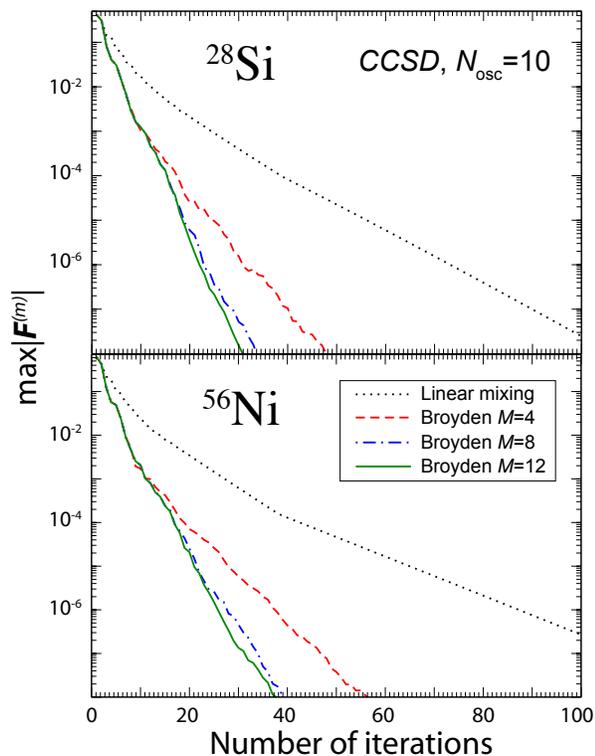}
\caption{(Color online) Convergence of \CCSD\ ground state energy of
$^{28}$Si (top) and $^{56}$Ni (bottom) using the modified Broyden's method
with $M$=4, 8, and 12 compared to the simple  mixing with  $\alpha$=0.7.
The calculations were carried out in a model space of $N_{\rm osc}$=10 major
oscillator shells.}
\label{ccfig1}
\end{figure}

Figure~\ref{ccfig1} compares the performance of the modified Broyden's method 
with $M$=4, 8, and 12
to the simple  mixing ansatz (\ref{LM}). To  
solve the \CCSD\ equations for the \CCSD\ ground state
energies of $^{28}$Si (top) and $^{56}$Ni (bottom),  we have used an oscillator
representation of the Hamiltonian in a model space of $N_{\rm osc}$=10 major
oscillator shells with $\hbar\omega_{\rm osc}$=32\,MeV. The Hamiltonian consists
of the kinetic-energy operator  (minus the center-of-mass correction)
 and the  N$^3$LO
nucleon-nucleon interaction of Ref.~\cite{EM02}.
It is seen that utilizing Broyden's  method increases the
convergence dramatically as compared to the simple mixing.  It is found
that for $M$=12, iterations converge more rapidly, and the tolerance of
10$^{-8}$  is reached typically after 30-40 iterations. For the linear
mixing to achieve similar accuracy, 130-140 iterations are required.

\section{Conclusions}\label{conclusions}
Broyden's method, 
a quasi-Newton method for the numerical solution of nonlinear equations 
in many variables, is
widely used in quantum
chemistry to perform  electronic-structure
calculations. In this work, we applied Broyden's method 
(both in its standard and modified variants)
to several  nuclear structure problems.
The examples range from  several \DFT\  applications
in various geometries to  {\it ab-initio} \CC\ calculations. 

The number of unknown variables (i.e., the size of the Broyden vector)
in our calculations ranges from $\sim$200 in {\SLDA} calculations with the
standard Broyden's method, to  $\sim$300,000 in {\HFBTHO}, $\sim$3,000,000
in {\HFODD}, and $\sim$10$^8$ in {\CC} applications of the modified 
Broyden's algorithm. Much faster convergence has been  achieved in
comparison with the linear-mixing procedure which is often used in such
types of calculations. We note that
Broyden's method often achieves superlinear convergence,
although mildly. 
If the iteration is not
properly structured, then it might not correspond to a contractive
mapping and may not converge. By
constructing an approximation to the Jacobian, Broyden's method can
determine the downhill direction even if the iteration itself does not
provide this information.

The downhill direction chosen by
Broyden's method is towards a fixed point, not necessarily towards a
minimum of the functional. Thus, the iterative process  might 
easily  converge to
a maximum or a saddle point even when the iteration would naturally
head towards a minimum. If the goal is to find the minimum, a combination of
the linear mixing and Broyden's method can be efficiently used,
as demonstrated in the example of constrained \HFB\ calculations of Fig.~\ref{fig6}.
However, the above feature of Broyden's method to converge to maxima
or saddle points 
can be taken advantage of if the theoretical
objective {\it is} to determine local maxima or
saddle points (e.g., when studying fission pathways).

In summary, Broyden's method is easy to incorporate into fixed-point
codes, and provides impressive performance improvements in nuclear
many-body calculations.

\acknowledgements
Useful discussions with Jorge Mor\'e and Jason Sarich are
gratefully acknowledged.
The UNEDF
SciDAC Collaboration is supported by the U.S. Department of Energy
under grant No. DE-FC02-07ER41457.
This work was also supported  by the U.S.~Department of Energy under
Contract Nos.~DE-FG02-96ER40963 (University of Tennessee),
DE-AC05-00OR22725 with UT-Battelle, LLC (Oak Ridge National
Laboratory), DE-FG05-87ER40361 (Joint Institute for Heavy Ion
Research), and DE-FG02-97ER41014 (University of Washington), 
by the National Nuclear Security
Administration under the Stewardship Science Academic Alliances 
program through the U.S. Department of Energy Research Grant 
DE-FG03-03NA00083, and by
the Polish Ministry of Science and Education under Contract N 202
179 31/3920. WN acknowledges support from
the  Carnegie Trust during his stay in Scotland.


\begin{thebibliography}{99}

%
% Introduction
%
\bibitem{RIAtheory}
{\it RIA Theory Bluebook: A Road Map},\\
http://www.orau.org/ria/RIATG/Blue\_Book\_FINAL.pdf.

\bibitem{unedf}
http://unedf.org.

\bibitem{KS}
W. Kohn and L.J. Sham, Phys. Rev. A {\bf 140}, 1133 (1965).


\bibitem{DFTpair}
L.N. Oliveira, E.K.U. Gross, and W. Kohn, Phys. Rev.
Lett. {\bf 60}, 2430 (1988);
N.N. Lathiotakis, M.A.L. Marques, M. L\"uders, L. Fast,
and E.K.U. Gross, Int. J. Quant. Chem. {\bf 99}, 790 (2004);
M. L\"uders, M.A.L. Marques, N.N. Lathiotakis, A. Floris,
G. Profeta, L. Fast, A. Continenza, S. Massidda, and
E.K.U. Gross, Phys. Rev. B {\bf 72}, 024545 (2005).

\bibitem{DFTrev}
M. Bender, P.-H. Heenen, and P.-G. Reinhard, Rev. Mod. Phys. {\bf 75}, 121 (2003);
{\it Extended Density Functionals in Nuclear Structure Physics}, ed. by G.A.
Lalazissis, P. Ring, and D. Vretenar (Springer Verlag, 2004).

\bibitem{LDA}
	E. Perli\'nska, S. G. Rohozi\'nski, J. Dobaczewski, and W. Nazarewicz,
	Phys. Rev. {\bf C69}, 014316 (2004).

\bibitem{broyden}
  C.G. Broyden, Math. Comput. {\bf 19}, 577 (1965).
\bibitem{global_broyden}
  J.E. Dennis and R.B. Schnabel, {\it Numerical Methods for Unconstrained Optimization and
  Nonlinear Equations}, (Prentice-Hall, 1983);
  W.H. Press, S.A.~Teukolsky, W.T. Vetterling,  and B.P. Flannery,
  {\it Numerical Recipies: The Art of Scientific Computing}, Third
  ed. (Cambridge University Press, 2007).

\bibitem{vanderbilt} 
D. Vanderbilt and S. G. Louie, Phys. Rev. B {\bf 30}, 6118 (1984).

\bibitem{mbm}
 D.D. Johnson, Phys. Rev. B {\bf 38}, 12807 (1988). 
 
 \bibitem{Eyert}
V. Eyert, J. Comput. Phys. {\bf 124}, 271 (1996).  

\bibitem{pwscf}
{Plane-Wave Self-Consistent Field (\PWscf) computer code, http://www.pwscf.org }

%
% Section II
%
\bibitem{unitaryGas}
	{
	A. Bulgac, Phys. Rev. A {\bf 76}, 040502(R) (2007);
	A. Bulgac and M. M. Forbes, arXiv:0804.3364;
	A. Bulgac and M. M. Forbes, {\em In preparation}, (2008)
	}
\bibitem{Randeria}	
	R. Sensarma, W. Schneider, R.B. Diener, and M. Randeria,
	arXiv:0706.1741 (2007).
\bibitem{DVR-1}
	{
	R.G. Littlejohn and M. Cargo,
	J. Chem. Phys. {\bf 117}, 27 (2002).
	}
\bibitem{DVR-2}
	{
	R.G. Littlejohn, M. Cargo, J. Tucker Carrington, K.A. Mitchell, and B. Poirier,
	J. Chem. Phys. {\bf 116}, 8691 (2002).
	}
\bibitem{DVR-3}
	{
	D. Baye, Phys. Stat. Sol. {\bf 243}, 1095 (2006).
	}

%
% Section III
%

\bibitem{hfodd}
	J. Dobaczewski and J. Dudek, 
	Comput. Phys. Commun. {\bf 102}, 166 (1997).
\bibitem{hfodd-1}
	J. Dobaczewski and J. Dudek, 
	Comput. Phys. Commun.{\bf 102}, 183 (1997);
	J. Dobaczewski and J. Dudek, 
	Comput. Phys. Commun. {\bf 131}, 164 (2000);
	J. Dobaczewski and P. Olbratowski, 
	Comput. Phys. Commun. {\bf 158}, 158 (2004);
	J. Dobaczewski and P. Olbratowski, 
	Comput. Phys. Commun.{\bf 167}, 214 (2005).

\bibitem{Dob96}
J. Dobaczewski, W. Nazarewicz,
T.R. Werner, J.-F. Berger, C.R. Chinn, and J. Decharg\'e,
Phys. Rev. {\bf C53}, 2809 (1996).	
\bibitem{Dob02} J. Dobaczewski, W. Nazarewicz, and  M.+V. Stoitsov,
Eur. Phys. J. A {\bf 15}, 21 (2002)
\bibitem{Bor06}
P.J. Borycki, J. Dobaczewski, W. Nazarewicz, and M.V. Stoitsov
Phys. Rev. {\bf C73}, 044319 (2006).
\bibitem{Bul02}
A. Bulgac, Phys. Rev. C {\bf 65}, 051305(R) (2002).

\bibitem{SkM}
J. Bartel, P. Quentin, M. Brack, C. Guet, and H.B.
           H{\aa}kansson, Nucl. Phys. {\bf A386}, 79 (1982).
\bibitem{SkL}    
E. Chabanat, P. Bonche, P. Haensel, J. Meyer,
and R. Schaeffer, Nucl. Phys. {\bf A627}, 710 (1997); 
{\bf A635}, 231 (1998).
     
\bibitem{SkP} 
J. Dobaczewski, H. Flocard,  and J. Treiner, Nucl. Phys.
{\bf A422}, 103 (1984).
           
\bibitem{Sto07}
M.V. Stoitsov, J. Dobaczewski, R. Kirchner,
W. Nazarewicz, and J. Terasaki,
Phys. Rev. C {\bf 76}, 014308 (2007).
%
% Section IV
%

\bibitem{dftresults}
 M.V. Stoitsov, J. Dobaczewski, W. Nazarewicz, S. Pittel, and D.J. Dean,
Phys. Rev. C {\bf 68}, 054312 (2003).

\bibitem{hfbtho}
M.V. Stoitsov, J. Dobaczewski, W. Nazarewicz, and P. Ring,
Comput. Phys. Commun. {\bf 167}, 43 (2005).

\bibitem{Stob} M.V. Stoitsov, in {\it Nuclear Theory}, ed. by S. Dimitrova,
Proc. 26th Int. Workshop on Nuclear Theory, Rila Mountains, Bulgaria, June 2007
(Institute for Nuclear Research and Nuclear Energy Bulgarian Academy of Sciences, Sofia,
2008), p. 13.

%
% Section V
%
\bibitem{Coe58}
F.\ Coester, Nucl.\ Phys.\ {\bf 7}, 421 (1958).
\bibitem{Coe60}
F.\ Coester and H.\ K{\" u}mmel, Nucl.\ Phys.\ {\bf 17}, 477 (1960).
\bibitem{Bar07}
R.J.\ Bartlett and M.\ Musia{\l}, \rmp {\bf 79}, 291 (2007).
\bibitem{Dea05}
D.J. Dean J.R. Gour, G. Hagen, M. Hjorth-Jensen, K. Kowalski, 
T. Papenbrock, P. Piecuch, and M. W{\l}och,
Nucl. Phys. A {\bf 752}, 299 (2005).
\bibitem{Hag06}
G.\ Hagen, D.J.\ Dean, M.\ Hjorth-Jensen, and T.\ Papenbrock,
Phys. Lett. B {\bf 656}, 169 (2007).
\bibitem{Hag07a}
G.\ Hagen, T.\ Papenbrock, D.J.\ Dean, A.\ Schwenk, A.\ Nogga, 
 M.\ W{\l}loch, and P.\ Piecuch 
\prc {\bf 76}, 034302 (2007).
\bibitem{Hag07b}
G.\ Hagen, D.~J.\ Dean, M.\ Hjorth-Jensen, T.\ Papenbrock, A.\
Schwenk, 
\prc {\bf 76}, 044305 (2007).
\bibitem{gour06}
J.R. Gour, P. Piecuch, M. Hjorth-Jensen, M. W{\l}och, and D.J. Dean, 
\prc {\bf 74}, 024310 (2006).
\bibitem{EM02}
D.R. Entem and R. Machleidt, Phys. Lett. B {\bf 524}, 93
(2002).

	
\end{thebibliography}
\end{document}